\documentclass[twoside,psfig,10pt]{article}
\usepackage{amsmath}

\title{ 
\Large
{\bf On pulse broadening for optical solitons} 
}
\author{ 
{\bf H.J.S. Dorren} \; \;  
{\bf and} \; \;
{\bf J.J.B. van den Heuvel}
\\
{\small Department of Electrical Engineering, 
Eindhoven University of Technology,
P.O. Box 513,} \\
{\small 5600 MB Eindhoven, The Netherlands} }
\date{} 

\begin{document}
\maketitle

\begin{abstract}
Pulse broadening for optical solitons due to birefringence is
investigated. We present an analytical solution which 
describes the propagation of solitons in birefringent optical fibers.
The special solutions consist of 
a combination of purely solitonic terms propagating along the principal 
birefringence axes and soliton-soliton interaction terms. 
The solitonic part of the solutions indicates that 
the decay of initially localized pulses could be due 
to different propagation velocities along the birefringence 
axes. We show that the disintegration of solitonic pulses
in birefringent optical fibers can be caused by two effects. 
The first effect is similar as in linear birefringence and is related
to the unequal propagation velocities of the modes along the birefringence
axes. The second effect is
related to the nonlinear soliton-soliton interaction between
the modes, which makes 
the solitonic pulse-shape blurred. 
%\newline
%PACS-numbers: 42.65.-k, 42.81.Dp, 02.30.Jr
\end{abstract}

\vspace{3cm}
\begin{flushleft}
\begin{small}
\hspace{1cm}
E-mail: H.J.S.Dorren@ele.tue.nl
\end{small}
\end{flushleft}

\newpage
\section{Introduction}
The demand for long-distance high-bandwidth data transfer in the field
of optical telecommunications has lead to research about the use of
optical envelope solitons as information carriers. Solitons are ideal 
for optical telecommunications because the dispersion of the optical
fiber is exactly counterbalanced by the nonlinearity. As a result of this,
the information carrier can maintain its pulse-shape over long distances.
In an ideal optical fiber, the nonlinear Schr\"{o}dinger
equation describes the electromagnetic field envelope in a single
polarization case. Actual optical fibers however, have the property
that there is a difference in the propagation velocity for the two
different polarization states of the electromagnetic field. This
property is called birefringence. The effects of birefringence are almost
never steady, but they vary randomly, in both magnitude and orientation.
The random birefringence will ensure that an initially localized pulse will 
eventually disintegrate. This effect 
is called Polarization Mode Dispersion (PMD). PMD is important in situations
where high-bit-rate optical signals have to be transported over long
distances such as sub-marine intercontinental optical connections. This is
the reason that the  topic has already had considerable attention in both
theoretical and experimental research in the field of optical 
telecommunications \cite{Fontaine}-\cite{Gisin}. 

The dynamics of nonlinear
waves in birefringent optical fibers is described in the literature
by the following set of coupled nonlinear differential equations, which were
originally introduced by Berkhoer and Zakharov \cite{Berkhoer}:
\begin{equation}
\begin{array}{l}
\displaystyle
i u_{1x} +  i \delta u_{1t} + \frac{1}{2} u_{1tt} = -(|u_{1}|^2 +\gamma |u_{2}|^2) u_{1},  \\
\displaystyle
i u_{2x} -  i \delta u_{2t} + \frac{1}{2} u_{2tt} = -(|u_{2}|^2 +\gamma |u_{1}|^2) u_{2}.
\end{array}
\label{eq: manakov}
\end{equation}
A detailed derivation of Eq.(\ref{eq: manakov}), which are from now on called
the birefringence equations, can be found in the book by Hasegawa \cite{Hasegawa}.
In Eq.(\ref{eq: manakov}), the (scaled) electromagnetic field envelopes 
of the different polarization states (modes) are described by $u_{1}$ and $u_{2}$ 
respectively. The parameter $\gamma$ describes the strength of the cross-phase 
modulation. For single mode optical fibers we can use $\gamma = \frac{2}{3}$. 
The parameter $\delta$ describes the group
velocity birefringence. In the limit $\delta \rightarrow 0$, 
the well known Manakov equation is retained. 
Ueda and Kath have argued that the parameter 
$\delta$ is actually a random parameter with zero mean \cite{Ueda}:
\begin{equation}
\langle \delta \rangle = 0.
\end{equation}
As a result of this, Eq.(\ref{eq: manakov}) is a stochastic differential
equation. In the following, we treat the parameter $\delta$  
as the half-difference of the group velocity 
between the local principal birefringence axes.
If we are able to solve Eq.(\ref{eq: manakov}) analytically however,
we can obtain analytical expressions for the propagation properties of an 
initially localized soliton in a birefringent medium. 

The set of coupled equations (\ref{eq: manakov}) have also been investigated 
by others. In Ref.\cite{Menyuk} the behavior of solutions of Eq.(\ref{eq: manakov})
is investigated numerically. One of the conclusions of Ref.\cite{Menyuk} is that 
for solitonic initial conditions the two partial pulses (modes) always move together. 
A complicated pattern of soliton-soliton interactions between the modes
is presented. Another important 
result is presented in Ref.\cite{Kath}. Firstly, it is shown in this
publication that by applying the transformation:
\begin{equation}
\begin{array}{l}
\displaystyle
u_{1} = \tilde{u}_{1} \exp \left[ i \frac{ \delta^{2} }{2} x - i \delta t \right], \\
\displaystyle
u_{2} = \tilde{u}_{2} \exp \left[ i \frac{ \delta^{2} }{2} x + i \delta t \right],
\end{array}
\label{trans}
\end{equation}
the set of equations Eq.(\ref{eq: manakov}) transform into:
\begin{equation}
\begin{array}{l}
\displaystyle
i \tilde{u}_{1x} + \frac{1}{2} \tilde{u}_{1tt} = -(|\tilde{u}_{1}|^2 
+\gamma |\tilde{u}_{2}|^2) \tilde{u}_{1},  \\
\displaystyle
i \tilde{u}_{2x} + \frac{1}{2} \tilde{u}_{2tt} = -(|\tilde{u}_{2}|^2 
+\gamma |\tilde{u}_{1}|^2) \tilde{u}_{2}.
\end{array}
\label{eq: manakov0}
\end{equation}
If $\gamma=1$, the set of equations 
(\ref{eq: manakov0}) are called the Manakov equations. The Manakov
equations are well-studied and solutions can be obtained in several 
ways \cite{Manakov,Zakharov,Radh,Shches,Zen1,Zen2,Zen3}. 
It has to be remarked that the 
transformation (\ref{trans}) leads to special solutions of 
Eq.(\ref{eq: manakov}), in which the effect of the birefringence appears 
in the phase factor.
Since under experimental conditions usually the optical power is measured
({\em i.e.} $|u_{i}|^{2}$), the transformation 
(\ref{trans}) leads to special solutions of Eq.(\ref{eq: manakov}),
which can not explain the decay of optical pulses due to 
birefringence (since the birefringence is in the phase). 
On the other 
hand in Ref.\cite{Kath} solutions are reported which show that 
Eq.(\ref{eq: manakov0}) has modes which propagate with a different 
velocity. A general method to solve the equations (\ref{eq: manakov})  
has never been given \cite{Hasegawa}.

There are more approaches possible which can be used to find solutions 
of Eq.(\ref{eq: manakov}). Although Hirota's method has been proved 
to be successful for many types of equations, we approach the problem
in this paper using a method developed by one of us (see Ref.\cite{Dorren})
to find special solutions of Eq.(\ref{eq: manakov}). These solutions
have the 
property that the solitonic 
part of the modes propagate with a different velocity. The existence of 
such a solution might be important for explaining the decay of 
solitonic pulses in birefringent optical fibers. For linear birefringence
it is well known that the decay of initially localized pulses can be 
explained by the fact that both modes propagate with a different 
velocity along the birefringence axes 
\cite{Agrawal}. The results of this paper suggests that for solitonic
pulses a similar behavior takes place.

In Sect.II, we apply concepts derived in earlier publications to find 
special solutions of the birefringence problem.  
We derive solutions consisting of 
of two unperturbed solitons propagating with a different
velocity along the principal birefringence axes. Moreover the solutions 
contain interactions terms between the pure solitonic solutions. 
It is concluded that two effects play a role in the decay of solitonic 
pulses. The effect is similar as in linear birefringence and is due to 
the unequal propagation velocity of the modes. The second effect, which 
is absent in linear birefringence is the nonlinear soliton-soliton
interaction between the modes.
The paper is concluded with a discussion.
%
% ************** NEW SECTION ********************
%
\section{Special solutions of the birefringence equations}
For solving Eq.(\ref{eq: manakov}), we try 
a starting point the following free wave solutions:
\begin{equation}
\begin{array}{l}
u_{1}(x,t)=A e^{i ( k_{1} t - \omega_{1} x)}, \\
u_{2}(x,t)=B e^{i ( k_{2} t - \omega_{2} x)}.
\end{array} 
\label{freewave}
\end{equation} 
If we substitute Eq.(\ref{freewave}) into 
Eq.(\ref{eq: manakov}), we obtain:
\begin{equation}
\begin{array}{l}
\left[ \omega_{1}  - \delta k_{1} - \frac{1}{2} k^{2}_{1} \right] 
A e^{i ( k_{1} t - \omega_{1} x)} = -
A |A|^{2} e^{2i(k_{1}t - \omega_{1}x)} 
e^{-i(\overline{k}_{1}t - \overline{\omega}_{1}x)} -
\gamma A |B|^{2} e^{i(k_{1}t - \omega_{1}x)} 
e^{i([k_{2} - \overline{k}_{2}]t - [\omega_{2} - \overline{\omega}_{2}]x)},
\\
\left[
\omega_{2}  + \delta k_{2} - \frac{1}{2} k^{2}_{2}
\right] B e^{i ( k_{2} t - \omega_{2} x)} = -
B |B|^{2} e^{2i(k_{2}t - \omega_{2}x)}
e^{-i(\overline{k}_{2}t - \overline{\omega}_{2}x)} -
\gamma B |A|^{2} e^{i(k_{2}t - \omega_{2}x)}
e^{i([k_{1} - \overline{k}_{1}]t - [\omega_{1} - \overline{\omega}_{1}]x)}.
\\
\end{array}
\label{f_check}
\end{equation} 
In Eq.(\ref{f_check}), it is assumed that $k_{i}$ and $\omega_{i}$ are both
complex (the bars indicate the complex conjugates). 
We require that the following dispersion relationships are valid: 
\begin{equation}
\begin{split}
\omega_{1}  &= + \delta k_{1} + \frac{1}{2} k^{2}_{1},  \\
\omega_{2}  &= - \delta k_{2} + \frac{1}{2} k^{2}_{2}.  
\end{split}
\label{dispersion}
\end{equation}
We can conclude that for nonzero coefficients $A$ and $B$, the try-solution
(\ref{freewave}) does not satisfy Eq.(\ref{eq: manakov}) and hence the  
birefringence equations have no free wave solutions.
Before deriving special solutions of Eq.(\ref{eq: manakov}) the following
simplifications are imposed to simplify the 
computations which will follow:
\begin{equation}
\begin{split}
I&:   \ \ \   k_{1}=k_{2}=k;  \hspace{0.5cm} k=a+ib, \\
II&:  \ \ \   a=0;            \hspace{0.5cm} k=ib. \\
\end{split}
\end{equation}
Condition I is related to the fact that we strive for a situation 
in which both the linear part as the nonlinear part of Eq.(\ref{eq: manakov}) 
are expanded in the same basis functions. 
In experiments, this situation can be obtained by
optimizing the ``launching conditions''. Condition II is introduced
by realizing that in fiber optics only envelope solutions can 
be measured. 
In this paper, we search for modes $u_{1}$ and $u_{2}$ propagating with 
unequal velocity. We therefore substitute the following 
solutions into Eq.(\ref{eq: manakov}):
\begin{equation}
\begin{split}
u_{1}(x,t) &= e^{\frac{1}{2}ib^{2} x} 
\sum_{n=0}^{\infty} \sum_{m=0}^{\infty} 
\hat{A}_{2n+1,m} e^{-[2n+1]bz_{1}} e^{-2mbz_{2}}, \\
u_{2}(x,t) &= e^{\frac{1}{2} i b^{2} x}
\sum_{n=0}^{\infty} \sum_{m=0}^{\infty}
\hat{B}_{2n+1,m} e^{-2mbz_{1}} e^{-[2n+1]bz_{2}}. 
\end{split}
\label{soliton}
\end{equation}
In Eq.(\ref{soliton}) it is used that $z_{1}=t - \delta x$ and 
$z_{2}=t + \delta x$. For the solutions (\ref{soliton}) the left-hand
side of Eq.(\ref{eq: manakov}) is equal to:
\begin{equation}
\begin{array}{l}
\displaystyle
i u_{1x} +  i \delta u_{1t} + \frac{1}{2} u_{1tt}=
e^{\frac{1}{2}ib^{2} x}
\sum_{n=0}^{\infty} \sum_{m=0}^{\infty}
\left[
- \frac{1}{2} b^{2} - 4 m i b \delta + \frac{1}{2} \left( 2n + 2m  + 1 \right)^{2} b^{2}
\right]
\hat{A}_{2n+1,m}  e^{-[2n+1]bz_{1}} e^{-2mbz_{2}}, \\
\displaystyle
i u_{2x} -  i \delta u_{2t} + \frac{1}{2} u_{2tt}=
e^{\frac{1}{2}ib^{2} x}
\sum_{n=0}^{\infty} \sum_{m=0}^{\infty}
\left[
- \frac{1}{2} b^{2} + 4 m i b \delta + \frac{1}{2} \left( 2n + 2m  + 1 \right)^{2} b^{2}
\right]
\hat{B}_{2n+1,m} e^{-[2n+1]bz_{2}} e^{-2mbz_{1}}.
\end{array} 
\label{leftt}
\end{equation} 
We can conclude that the linear part of Eq.(\ref{eq: manakov}) is not 
modifying the structure of the exponential functions (\ref{soliton}).
This appears also to be the case for the nonlinear part. 
If we substitute Eq.(\ref{soliton}) into the right-hand side of 
Eq.(\ref{eq: manakov}) we obtain: 
\begin{equation}
\begin{split}
| u_{1}^{2} | u_{1} + \gamma | u_{2}^{2} | u_{1} &=
e^{\frac{1}{2} i b^{2} x }
\sum_{n=1}^{\infty}
\sum_{m=0}^{\infty}
\sum_{p=0}^{n-1}
\sum_{q=0}^{p}
\sum_{v=0}^{m}
\sum_{w=0}^{v}
\hat{A}_{2q+1,w} \hat{A}_{2p-2q+1,v-w} \hat{A}_{2n-2p-1,m-v}
e^{-[2n+1]bz_{1}} e^{-2mbz_{2}} \\
&+ 
\gamma
e^{ \frac{1}{2} i b^{2} x }
\sum_{n=1}^{\infty}
\sum_{m=1}^{\infty}
\sum_{p=0}^{n}
\sum_{q=0}^{p}
\sum_{v=0}^{m-1}
\sum_{w=0}^{v}
\hat{B}_{2m-2v-1,n-p} \hat{B}_{2w-1,q} \hat{A}_{2p-2q+1,v-w}
e^{-[2n+1]bz_{1}} e^{-2mbz_{2}}, \\
| u_{2}^{2} | u_{2} + \gamma | u_{1}^{2} | u_{2} &=
e^{\frac{1}{2} i b^{2} x }
\sum_{n=0}^{\infty}
\sum_{m=0}^{\infty}
\sum_{p=0}^{n-1}
\sum_{q=0}^{p}
\sum_{v=0}^{m}
\sum_{w=0}^{v}
\hat{B}_{2q+1,w} \hat{B}_{2p-2q+1,v-w} \hat{B}_{2n-2p-1,m-v}
e^{-[2n+1]bz_{2}} e^{-2mbz_{1}}  \\
&+
\gamma
e^{\frac{1}{2} i b^{2} x }
\sum_{n=1}^{\infty}
\sum_{m=1}^{\infty}
\sum_{p=0}^{n}
\sum_{q=0}^{p}
\sum_{v=0}^{m-1}
\sum_{w=0}^{v}
\hat{A}_{2m-2v-1,n-p} \hat{A}_{2w-1,q} \hat{B}_{2p-2q+1,v-w}
e^{-[2n+1]bz_{2}} e^{-2mbz_{1}}.
\end{split}
\label{rightt}
\end{equation}
It can be concluded by comparing Eq.(\ref{leftt}) and Eq.(\ref{rightt}) that  
solutions of the type (\ref{soliton}) ensure that both the linear part and the nonlinear
part of Eq.(\ref{eq: manakov}) can be expanded in the same basis functions.
Given non-zero coefficients 
$\hat{A}_{1,0}$ and $\hat{B}_{1,0}$ ($A,B \in I\!\!R$), all the other coefficients $\hat{A}_{2n+1,m}$ and 
$\hat{B}_{2n+1,m}$ of the solution (\ref{soliton}) are determined by the following iteration series: 
\begin{equation}
\begin{split}
\displaystyle
\hat{A}_{2n+1,m} = -
&\sum_{p=0}^{n-1}
\sum_{q=0}^{p}
\sum_{v=0}^{m}
\sum_{w=0}^{v}
\frac{
\hat{A}_{2q+1,w} \hat{A}_{2p-2q+1,v-w} \hat{A}_{2n-2p-1,m-v} 
}{
- \frac{1}{2} b^{2} - 4 m i b \delta + \frac{1}{2} \left( 2n + 2m  + 1 \right)^{2} b^{2}
} 
\\ 
- 
&\gamma \sum_{p=0}^{n}
\sum_{q=0}^{p}
\sum_{v=0}^{m-1}
\sum_{w=0}^{v}
\frac{
\hat{B}_{2m-2v-1,n-p} \hat{B}_{2w-1,q} \hat{A}_{2p-2q+1,v-w}
}{
- \frac{1}{2} b^{2} - 4 m i b \delta + \frac{1}{2} \left( 2n + 2m  + 1 \right)^{2} b^{2}
},
\\
\displaystyle
\hat{B}_{2n+1,m} = -
&\sum_{p=0}^{n-1}
\sum_{q=0}^{p}
\sum_{v=0}^{m}
\sum_{w=0}^{v}
\frac{
\hat{B}_{2q+1,w} \hat{B}_{2p-2q+1,v-w} \hat{B}_{2n-2p-1,m-v}
}{
- \frac{1}{2} b^{2} + 4 m i b \delta + \frac{1}{2} \left( 2n + 2m  + 1 \right)^{2} b^{2}
}
\\ 
-
& \gamma \sum_{p=0}^{n}
\sum_{q=0}^{p}
\sum_{v=0}^{m-1}
\sum_{w=0}^{v}
\frac{
\hat{A}_{2m-2v-1,n-p} \hat{A}_{2w-1,q} \hat{B}_{2p-2q+1,v-w}
}{
- \frac{1}{2} b^{2} + 4 m i b \delta + \frac{1}{2} \left( 2n + 2m  + 1 \right)^{2} b^{2}
}, 
\end{split}
\label{recur}
\end{equation}
where in the second terms on the right-hand side $m>0$.

The solution (\ref{soliton}) of Eq.(\ref{eq: manakov}) is purely formal and
does not give adequate insight of the behavior of solitonic solutions 
in birefringent media. In order to obtain this insight, it is convenient
to separate the solutions $u_{1}(x,t)$ and $u_{2}(x,t)$ into two parts:
\begin{equation}
\begin{split}
u_{1}(x,t) &=
e^{\frac{1}{2}ib^{2} x}
\sum_{n=0}^{\infty}
\hat{A}_{2n+1,0} e^{-[2n+1]bz_{1}} +
e^{\frac{1}{2}ib^{2} x} 
\sum_{n=0}^{\infty} \sum_{m=1}^{\infty} 
\hat{A}_{2n+1,m} e^{-[2n+1]bz_{1}} e^{-2mbz_{2}}, \\
u_{2}(x,t) &=
e^{\frac{1}{2}ib^{2} x}
\sum_{n=0}^{\infty}
\hat{B}_{2n+1,0} e^{-[2n+1]bz_{2}} +
e^{\frac{1}{2}ib^{2} x}
\sum_{n=0}^{\infty} \sum_{m = 1}^{\infty}
\hat{B}_{2n+1,m} e^{-[2n+1]bz_{2}} e^{-2mbz_{1}}.
\end{split}
\label{separ}
\end{equation}
In this special case, we find that given  the coefficients $\hat{A}_{2n+1,0}$ and
$\hat{B}_{2n+1,0}$ are given by:
\begin{equation}
\begin{split}
\displaystyle
\hat{A}_{2n+1,0} = -
&\sum_{p=0}^{n-1}
\sum_{q=0}^{p}
\frac{
\hat{A}_{2q+1,0} \hat{A}_{2p-2q+1,0} \hat{A}_{2n-2p-1,0} 
}{
- \frac{1}{2} b^{2} + \frac{1}{2} \left( 2n + 1 \right)^{2} b^{2}
} 
\\ 
\displaystyle
\hat{B}_{2n+1,0} = -
&\sum_{p=0}^{n-1}
\sum_{q=0}^{p}
\frac{
\hat{B}_{2q+1,0} \hat{B}_{2p-2q+1,0} \hat{B}_{2n-2p-1,0}
}{
- \frac{1}{2} b^{2} + \frac{1}{2} \left( 2n + 1 \right)^{2} b^{2}
}
\\ 
\end{split}
\label{recur1}
\end{equation}
We firstly analyze the first part the right-hand side of Eq.(\ref{separ}). 
By assuming that $\hat{A}_{1,0}=A$ and $\hat{B}_{1,0}=B$ all the other 
coefficients $\hat{A}_{2n+1,0}$ and $\hat{B}_{2n+1,0}$ follow from Eq.(\ref{recur1}). 
This implies that Eq.(\ref{separ}) can be reformulated in 
the following form \cite{Dorren1}:
\begin{equation}
\begin{array}{l}
\displaystyle
u_{1}(x,t) = \frac{1}{2} A e^{\frac{1}{2} i b^{2} x} e^{i\xi_{0}}
\mbox{sech}(b[t - \delta x] + \xi_{0} ) 
+
e^{\frac{1}{2}ib^{2} t}
\sum_{n=0}^{\infty} \sum_{m=1}^{\infty}
\hat{A}_{2n+1,m} e^{-[2n+1]bz_{1}} e^{-2mbz_{2}}, 
\hspace{1cm}
\xi_{0} = - \frac{1}{2}
\log
\left( 
\frac{A^{2}}{4 b^{2} }
\right);
\\
\displaystyle
u_{2}(x,t) = \frac{1}{2} B e^{\frac{1}{2} i b^{2} x} e^{i\hat{\xi}_{0}}
\mbox{sech}(b[t + \delta x] + \hat{\xi}_{0} ) 
+
e^{\frac{1}{2}ib^{2} t}
\sum_{n=0}^{\infty} \sum_{m=1}^{\infty}
\hat{B}_{2n+1,m} e^{-[2n+1]bz_{2}} e^{-2mbz_{1}},
\hspace{1cm}
\hat{\xi}_{0} = - \frac{1}{2}
\log
\left( 
\frac{B^{2}}{4 b^{2} } 
\right).
\\ 
\end{array}
\label{final}
\end{equation}
The first term on the right-hand side of Eq.(\ref{final}) represents
unperturbed solitons. 
The propagation velocity of these solitons is determined by the birefringence
coefficients $\delta$. It follows from Eq.(\ref{final}) that the two 
solitons $u_{1}(x,t)$ and $u_{2}(x,t)$ propagate with a relative velocity 
$2 b \delta$. 
This makes that even in an ``ideal situation'', in which the non-solitonic
terms in Eq.(\ref{final}) are negligible an initially localized pulse is 
still unstable due do the unequal propagation velocity of the solitonic
solutions along the principal birefringence axes.

The second effect, which leads to instabilities, is the interaction between 
the unperturbed solitonic solutions. 
The asymptotic expansion of these instabilities is given by the sum of  
terms on the right-hand side of Eq.(\ref{final}). In principle, by using the 
recursion relationship (\ref{recur}), we can compute all  
the expansion coefficients $\hat{A}_{2n+1,m}$ and $\hat{B}_{2n+1,m}$. 
Only in special cases it is possible to carry out the summation explicitly 
to obtain an analytical expression for the interaction.

In Appendix A it is indicated that the summation in Eq.(\ref{final})
converges if $\delta < b$.  The interaction terms are proportional to a factor depending on 
$b, \delta$ and $\gamma$. From this result the conclusion can be drawn 
that the interaction terms are small if (Appendix A):
\begin{equation}
\left|
\frac{
b(1 + \gamma)
}{
2 \delta - 6b
}
\right|
\ll 1. 
\label{yyy}
\end{equation}
If the condition (\ref{yyy}) is satisfied,  the interaction 
between the modes is small compared to the solitonic part, and 
hence, the only that causes initially localized pulses to disintegrate is the
unequal propagation velocity of the modes.

\section{Discussion}
We have shown that the equations (\ref{eq: manakov}) have special
solutions which consist of solitonic pulses each propagating with a 
different velocity along the principal birefringence axes. 
The purely solitonic pulses are subject to soliton-soliton interactions 
between the modes.
This indicates that there are two mechanisms in birefringent 
optical fibers which can be responsible for the decay of optical 
solitons in birefringent optical fibers.

The first mechanism is exactly equal as for linear optical 
pulses. The solitonic optical pulses can decay because of 
a different propagation velocity of the two modes along 
the principal birefringence axes. It is no surprise that
also solitonic pulses are subject to linear birefringence 
since the propagation
velocity of the solitonic part of the solutions (\ref{final}) 
is determined by the linear part (dispersion relations)
of the differential equation
(\ref{eq: manakov}). The nonlinearity in the differential 
equation (\ref{eq: manakov}) has only influence on the 
shape of the solutions and keeps the propagation velocity
unaffected. 
If we consider for instance
the propagation properties of solitons in birefringent optical fibers, 
we can conclude that for nonlinear birefringence similar criteria as for 
linear birefringence apply. 
If we remember that Eq.(\ref{eq: manakov}) 
is defined in a coordinate frame moving with velocity $v_{0}$, 
we find that solutions $u_{1}(x,t)$ and $u_{2}(x,t)$ propagate 
with velocities $v_{0} - ( \frac{\delta}{v_{0}} ) v_{0}$ 
and $v_{0} + ( \frac{\delta}{v_{0}} ) v_{0}$. 
In Ref.\cite{Gisin} it is shown by using a mathematical 
analysis based upon Ref.\cite{Kac} that in this case
the PMD in long birefringent fibers increases as the square 
root of the length. 

The second effect, which can lead to decay of solitonic
optical pulses is the nonlinear soliton-soliton 
interaction between the modes. This effect has been
reported in Refs.\cite{Menyuk,Kath}. 
In the numerical results presented in Ref.\cite{Menyuk}
it is concluded that only soliton-soliton interactions
played a role in the decay of solitons. It is not
observed that two initially solitonic solutions propagate
with a different velocity along the birefringence axes.
We think that this result can be explained by the 
fact that in Ref.\cite{Menyuk} perfect solitonic pulse-shapes are taken
as the initial condition. 
If we solve 
Eq.(\ref{eq: manakov0}), it follows from Eq.(\ref{dispersion}) that 
in the case that $\delta=0$, both modes can be chosen to
have equal dispersion relations. Hence the solutions 
of Eq.(\ref{eq: manakov0}) can be expanded in the following 
form:
\begin{equation}
\begin{split}
\tilde{u}_{1}(x,t) &= e^{\frac{1}{2}ib^{2} x} 
\sum_{n=1}^{\infty} 
\tilde{A}_{n} e^{-nbt}, \\
\tilde{u}_{2}(x,t) &= e^{\frac{1}{2}ib^{2} x}
\sum_{n=1}^{\infty} 
\tilde{B}_{n} e^{-nbt}. 
\end{split}
\label{soliton1}
\end{equation}
For nonzero coefficients $\tilde{A}_{1}$ and $\tilde{B}_{2}$, all the
other coefficients can be determined by a similar iteration series 
as presented in \cite{Dorren1}. This implies that the Manakov equations
(\ref{eq: manakov0}) have solitonic solutions.
It might be possible that by choosing purely solitonic initial conditions,
the numerical simulations in Ref.\cite{Menyuk} lead to the special 
solutions (\ref{trans}). 
In realistic (experimental) optical systems it is difficult
to make pure solitons and a certain amount of noise will always be present. 
The appearance of noise can cause that the modes propagate 
with a different velocity and that the actual situation is 
more complicated as presented in \cite{Menyuk}. 

We want to conclude by remarking that there is another
way to understand the result obtained in Eq.(\ref{final}). 
We therefore firstly implement the following Galileian 
transformations:
\begin{equation}
\begin{array}{l}
z_{1} = t - \delta x \\
z_{2} = t + \delta x 
\end{array}
\end{equation}
As a result of this transformation we find that for $\epsilon=1$,  
Eq.(\ref{eq: manakov}) can be reformulated as:
\begin{equation}
\begin{array}{l}
\displaystyle
i \hat{u}_{1x} +  \frac{1}{2} \hat{u}_{1z_{1}z_{1}} + |\hat{u}_{1}|^2 \hat{u}_{1} =
- \epsilon \gamma |\hat{u}_{2}|^2 \hat{u}_{1},  \\
\displaystyle
i \hat{u}_{2x} +  \frac{1}{2} \hat{u}_{2z_{2}z_{2}} + |\hat{u}_{2}|^2 \hat{u}_{2}  =
- \epsilon \gamma |\hat{u}_{1}|^2 \hat{u}_{2}.
\end{array}
\label{eq: manakovsc}
\end{equation}
Eq.(\ref{eq: manakovsc}) describes a perturbed nonlinear Schr\"{o}dinger equation
for both the modes $\hat{u}_{1}$ and $\hat{u}_{2}$. It should be remarked that both
the modes $\hat{u}_{1}$ and $\hat{u}_{2}$ have a different time-evolution due to 
the rescaled time. 
If we try to solve Eq.(\ref{eq: manakovsc}) by implementing a perturbation series:
\begin{equation}
\begin{array}{l}
\hat{u}_{1} = f_{0} + \epsilon f_{1} +  \epsilon^{2} f_{2} + \epsilon^{3} f_{3} + \cdots, \\
\hat{u}_{2} = g_{0} + \epsilon g_{1} +  \epsilon^{2} g_{2} + \epsilon^{3} g_{3} + \cdots, 
\end{array}
\end{equation}
we obtain for the lowest order:
\begin{equation}
\begin{array}{l}
i f_{0x} +  \frac{1}{2} f_{0z_{1}z_{1}} + |f_{0}|^2 f_{0} = 0, \\
i g_{0x} +  \frac{1}{2} g_{0z_{2}z_{2}} + |g_{0}|^2 g_{0} = 0.
\end{array}
\label{ss1}
\end{equation}
(\ref{eq: manakovsc}) is the nonlinear Schr\"{o}dinger equation.
The solutions for $f_{0}$ and $g_{0}$ are therefore 
nonlinear Schr\"{o}dinger solitons  which are defined  
with respect to the transformed time $z_{1} = t-\delta$ and 
$z_{2}= t+\delta$ respectively. 
Furthermore, $f_{0}$ and $g_{0}$ are contaminated 
with the higher order perturbations $f_{n}$ and $g_{n}$. 
This is exactly what we find in Eq.(\ref{final}) where the 
asymptotic behavior of $f_{n}$ and $g_{n}$ is presented. 
As expected, the functions $f_{n}$ and $g_{n}$ have a 
time-evolution which is a combination of (exponential) functions 
which are defined with respect to the transformed time
$z_{1}$ and $z_{2}$. 

\subsection*{Acknowledgments}
This research was supported by the Netherlands Organization for Scientific
Research (N.W.O.) through the ``N.R.C. Photonics'' grant.
%H.J.S. Dorren 
%acknowledges J. Hietarinta for making some constructive comments on the first 
%version of the manuscript.

\newpage
\section*{Appendix A}
\subsection*{Convergence of Eq.(\ref{final})}
\renewcommand{\theequation}{\mbox{A-\arabic{equation}}}
\setcounter{equation}{0}
In this Appendix the convergence of the summation in Eq.(\ref{final})
is discussed. We estimate the coefficients $\hat{A}_{2n+1,m}$ 
and $\hat{B}_{2n+1,m}$ with respect to $\hat{A}_{2n+1,0}$ and
$\hat{B}_{2n+1,0}$.  
\begin{equation}
\begin{split}
\displaystyle
\left | \hat{A}_{2n+1,m} \right|  &= 
\left|
\sum_{p=0}^{n-1}
\sum_{q=0}^{p}
\sum_{v=0}^{m}
\sum_{w=0}^{v}
\frac{
\hat{A}_{2q+1,w} \hat{A}_{2p-2q+1,v-w} \hat{A}_{2n-2p-1,m-v}
}{
- \frac{1}{2} b^{2} - 4 m i b \delta + \frac{1}{2} \left( 2n + 2m  + 1 \right)^{2} b^{2}
} \right. \\
&+
\left.
\gamma \sum_{p=0}^{n}
\sum_{q=0}^{p}
\sum_{v=0}^{m-1}
\sum_{w=0}^{v}
\frac{
\hat{B}_{2m-2v-1,n-p} \hat{B}_{2w-1,q} \hat{A}_{2p-2q+1,v-w}
}{
- \frac{1}{2} b^{2} - 4 m i b \delta + \frac{1}{2} \left( 2n + 2m  + 1 \right)^{2} b^{2}
} 
\right| 
\\
&\leq (1 + \gamma)
\left|
\sum_{p=0}^{n-1}
\sum_{q=0}^{p}
\sum_{v=0}^{m}
\sum_{w=0}^{v}
\frac{
\hat{C}_{2q+1,w} \hat{C}_{2p-2q+1,v-w} \hat{C}_{2n-2p-1,m-v}
}{
- \frac{1}{2} b^{2} - 4 m i b \delta + \frac{1}{2} \left( 2n + 2m  + 1 \right)^{2} b^{2}
} \right|_{ \hat{C}_{n,m} = \mbox{max} \{ \hat{A}_{n,m},\hat{B}_{n,m} \} } 
\! \! \! \! \! \! \! \! \!
\mbox{{\em (Step 1)}}
\\
&\leq (1 + \gamma)
\left(
\left|
\sum_{p=0}^{n-1}
\sum_{q=0}^{p}
\sum_{v=0}^{m}
\sum_{w=0}^{v}
\frac{
\hat{C}_{2q+1,w} \hat{C}_{2p-2q+1,v-w} \hat{C}_{2n-2p-1,m-v}
\left(
- \frac{1}{2} b^{2} + \frac{1}{2} \left( 2n + 2m  + 1 \right)^{2} b^{2}
\right)
}{
\left( - \frac{1}{2} b^{2} + \frac{1}{2} \left( 2n + 2m  + 1 \right)^{2} b^{2}
\right)^{2} 
+ 16 m^{2}  b^{2} \delta^{2}
}
\right| \right. \\
&+
\left.
\left|
\sum_{p=0}^{n-1}
\sum_{q=0}^{p}
\sum_{v=0}^{m}
\sum_{w=0}^{v}
\frac{
\hat{C}_{2q+1,w} \hat{C}_{2p-2q+1,v-w} \hat{C}_{2n-2p-1,m-v}
\left(
4 m  b \delta
\right)
}{
\left( - \frac{1}{2} b^{2} + \frac{1}{2} \left( 2n + 2m  + 1 \right)^{2} b^{2}
\right)^{2}
+ 16 m^{2}  b^{2} \delta^{2}
}
\right| \right)_{ \hat{C}_{n,m} = \mbox{max} \{ \hat{A}_{n,m},\hat{B}_{n,m} \} } 
\! \! \! \! \! \! \! \! \!
\mbox{{\em (Step 2)}}
\\
&\leq (1 + \gamma)
\left(
\left|
\sum_{p=0}^{n-1}
\sum_{q=0}^{p}
\sum_{v=0}^{m}
\sum_{w=0}^{v}
\frac{
\hat{C}_{2q+1,w} \hat{C}_{2p-2q+1,v-w} \hat{C}_{2n-2p-1,m-v}
\left(
- \frac{1}{2} b^{2} + \frac{1}{2} \left( 2n + 2m  + 1 \right)^{2} b^{2}
\right)
}{
\left( - \frac{1}{2} b^{2} + \frac{1}{2} \left( 2n + 2m  + 1 \right)^{2} b^{2}
\right)^{2} 
}
\right| \right. \\
&+
\left.
\left|
\sum_{p=0}^{n-1}
\sum_{q=0}^{p}
\sum_{v=0}^{m}
\sum_{w=0}^{v}
\frac{
\hat{C}_{2q+1,w} \hat{C}_{2p-2q+1,v-w} \hat{C}_{2n-2p-1,m-v}
\left(
4 m  b \delta
\right)
}{
\left( - \frac{1}{2} b^{2} + \frac{1}{2} \left( 2n + 2m  + 1 \right)^{2} b^{2}
\right)^{2}
}
\right| \right)_{ \hat{C}_{n,m} = \mbox{max} \{ \hat{A}_{n,m},\hat{B}_{n,m} \} } 
\! \! \! \! \! \! \! \! \!
\mbox{{\em (Step 3)}}
\\
&\leq 
\left|
\frac{
b(1+\gamma)
}{
%8 m \delta - 4 n^{2} b -4mb - 4nb - 8nmb 
2 \delta - 6 b
}
\right|
\left|
\sum_{p=0}^{n-1}
\sum_{q=0}^{p}
\frac{
\hat{C}_{2q+1,0} \hat{C}_{2p-2q+1,0} \hat{C}_{2n-2p-1,0}
}{
- \frac{1}{2} b^{2} + \frac{1}{2} \left( 2n + 1 \right)^{2} b^{2}
}
\right|_{ \hat{C}_{2n+1,0} = \mbox{max} \{ \hat{A}_{2n+1,0},\hat{B}_{2n+1,0} \} } 
\! \! \! \! \! \! \! \! \!
\mbox{{\em (Step 4)}}
\\
&=
\left|
\frac{
b(1+\gamma)
}{
2 \delta - 6 b  
}
\right|
\left|
\hat{C}_{2n+1,0}
\right|_{ \hat{C}_{2n+1,0} = \mbox{max} \{ \hat{A}_{2n+1,0},\hat{B}_{2n+1,0} \} }
\end{split}
\label{estimate}
\end{equation}
The coefficients $\hat{B}_{2n+1,m}$ can be estimated in a similar way as the coefficients 
$\hat{B}_{2n+1,m}$. 
In {\em Step 1}, we replace either the coefficients $\hat{A}_{n,m}$
or $\hat{B}_{n,m}$ by 
$\hat{C}_{n,m} = \mbox{max} \{ \hat{A}_{n,m},\hat{B}_{n,m} \}$. This
replacement is justified because:
\begin{equation}
| u_{1} |^{2} u_{1} + \gamma | u_{2} |^{2} u_{1}
\leq 
(1 + \gamma)  |v|^{2} v
\hspace{1cm}
v = \mbox{max} \{ u_{1},u_{2} \}
\end{equation}
In {\em Step 2}, the triangle inequality is applied to the complex 
denominator. In {\em Step 3}, is the quotient presented in in 
Eq.(\ref{estimate}) reaches an upper-limit by putting $m=0$ in 
the denominator.
Finally, in {\em Step 4} it is used that:
\begin{equation}
- \frac{1}{2} b^{2} + \frac{1}{2} \left( 2n + 2m  + 1 \right)^{2} b^{2} 
> 4mb \delta
\label{xxx}
\end{equation}
Eq.(\ref{xxx}) is valid if $\delta \leq b$ for all $n,m > 0$. 
Furthermore, the summation over $m$ is estimated to be smaller
than $m^{2}$ times the summation over coefficients for which $m=0$. 
From Eq.(\ref{xxx}), it follows that:
\begin{equation}
m^{2} < \frac{b}{2 \delta - 6 b}
\end{equation}
As a result of Eq.(\ref{estimate}) we can estimate an upper limit of the double sum 
in Eq.(\ref{final}). Because the coefficients $|\hat{A}_{2n+1,m}|$ and $|\hat{B}_{2n+1,m}|$
are proportional to $|\hat{A}_{2n+1,0}|$ and $|\hat{B}_{2n+1,0}|$, 
we can estimate the summation over $n$ in Eq.(\ref{final}) is proportional 
to the unperturbed soliton. The summation over $m$ can be carried out independently,
and remains finite because it forms a geometric series. Finally we obtain: 
\begin{equation}
\begin{array}{l}
\displaystyle
u_{1}(x,t) \leq  \frac{1}{2} A e^{\frac{1}{2} i b^{2} x} e^{i\xi_{0}}
\mbox{sech}(b[t - \delta x] + \xi_{0} ) 
+
\frac{1}{2} C
\left|
\frac{
b(1+\gamma)
}{
2 \delta - 6 b
}
%\right|
%\left|
%\mbox{sech}(b[t - \delta x]) 
%\mbox{sech}(bt)
\right|_{C = \mbox{max} \{ A,B \} },
\\
\displaystyle
u_{2}(x,t) = \frac{1}{2} B e^{\frac{1}{2} i b^{2} x} e^{i\hat{\xi}_{0}}
\mbox{sech}(b[t + \delta x] + \hat{\xi}_{0} ) 
+
\frac{1}{2} C
\left|
\frac{
b(1+\gamma)
}{
2 \delta - 6 b
}
%\right|
%\left|
%\mbox{sech}(b[t - \delta x])
%\mbox{sech}(bt)
\right|_{C = \mbox{max} \{ A,B \} }.
\\ 
\end{array}
\end{equation}

\end{document}